\documentclass[preprint,aps,amsmath]{revtex4}

\usepackage{graphicx}
\usepackage{subfig}
\usepackage{amssymb}
\usepackage{bm}

\renewcommand{\vec}[1]{\boldsymbol{\mathit{#1}}}
\newcommand\bnabla{\boldsymbol \nabla}

\begin{document}

\title{Gyrofluid computation of magnetic perturbation effects on
  turbulence and edge localized bursts}

\author{J. Peer$^1$, A. Kendl$^1$, T.T. Ribeiro$^2$ and B.D. Scott$^2$}

\affiliation{$^1$ Institut f\"ur Ionenphysik und Angewandte Physik, Universit\"at
  Innsbruck, Austria\\
$^2$ Max-Planck-Institut f\"ur Plasmaphysik, Garching, Germany \vspace{1cm}}

\begin{abstract}
\vspace{1cm}
    The effects of non-axisymmetric resonant magnetic perturbation fields
    (RMPs) on saturated drift-wave turbulence and on ballooning mode bursts in
    the edge pedestal of tokamak plasmas are investigated by numerical simulations
    with a nonlinear six-moment electromagnetic gyrofluid model including
    zonal profile evolution. 
    The vacuum RMP fields are screened by plasma response currents, so that 
    magnetic transport by perturbed parallel motion is not significantly changed. 
    Radial transport of both particles and heat is dominated by turbulent
    convection even for large RMP amplitudes, where formation of stationary
    convective structures leads to edge profile degradation.  
    Modelling of ideal ballooning mode unstable edge profiles for single bursts
    including RMP fields causes resonant mode locking and destabilization.
\end{abstract}


\maketitle

\section{Introduction}

In future tokamaks like ITER, the heat flux associated with edge-localised
modes (ELMs) is estimated to seriously damage the plasma facing components
(PFCs) \cite{federici03,federici06,loarte03}.  Thus, methods for the
suppression or at least effective mitigation of ELMs are essential for 
economically viable steady state operation of tokamaks. One of the most promising
techniques to control ELMs is the external application of non-axisymmetric
resonant and non-resonant magnetic perturbations which were found to increase
the ELM frequency and to reduce the heat load on the PFCs
\cite{evans04,evans05,evans06,liang07,liang10,kirk11,suttrop11a,suttrop11b}.
Models for the physics underlying the ELM mitigation by resonant magnetic
perturbation fields (RMPs) assume an
RMP-induced formation of ergodic magnetic field regions which modify the radial
transport and hence decrease the edge pressure gradient below the
peeling-ballooning instability threshold \cite{tokar07,tokar08,snyder07}. 

Numerical computations on the effects of RMPs face the problem that present
first-principles-based edge turbulence models are not able to obtain a
self-consistent H-mode edge transport barrier with realistically steep flow and
pressure profiles. Thus, a self-consistent treatment of the interaction between
ELMs and RMPs is not possible. In contrast, present simulation codes allow for
well-founded investigations on the effects of RMPs in L-mode edge turbulence conditions.
Previous computational studies on the effects of RMPs on turbulence where based
on two- and four-field reduced MHD models \cite{beyer02,becoulet12}
and four-field drift-fluid models \cite{reiser05,reiser07,reiser09}. 
The gyrofluid approach presented in this work allows to include
self-consistent electron and ion temperature dynamics. 
Thus, a more realistic treatment of the radial heat transport becomes possible.

Bearing in mind the consistency limitations concerning H-mode states, 
we then further examine two models for the edge transport barrier.  
As one approach, RMP fields are included in the standard (unmitigated)
``H-mode''-like profile model described in Refs.\ \cite{kendl10,peer13} for
gyrofluid computation of edge localized ideal ballooning mode bursts.
It is found that this standard profile scenario does not allow to reproduce direct
mitigation of ideal ballooning ELMs by RMPs. 

The work is organised as follows. After the introduction in sec.~1, the gyrofluid
computational model, the numerical setup and the method used to implement RMPs
are presented in sec.~2.  The turbulence computation results are evaluated and
analysed in sec.~3, and the ELM simulation methods and results are discussed
in sec.~4.   
Conlusions are given in sec.~5.

\section{Computational gyrofluid and geometry model including RMPs}

The computations presented are carried out with the nonlinear
gyrofluid electromagnetic model and code GEMR \cite{scott05}.  The model
includes six moment equations each for electrons and ions (labelled with
$z\in\{\mathrm{e},\mathrm{i}\})$ which are coupled by a polarisation equation
and an induction equation.  The dependent variables are density $n_{z}$,
parallel velocity $u_{z\parallel}$, parallel temperature $T_{z\parallel}$,
perpendicular temperature $T_{z\perp}$, parallel component of the parallel heat
flux $q_{z\parallel\parallel}$, perpendicular component of the parallel heat
flux $q_{z\parallel\perp}$, electric potential $\phi$ and parallel magnetic
potential $A_{\parallel}$. The full set of model equations is presented in
Refs.\ \cite{kendl10,scott05}.

The model uses normalised quantities, where the perpendicular spatial
scales are given in units of the minor plasma radius $a$. 
The time scale is normalised by $a/c_{\mathrm{s}0}$, where $c_{\mathrm{s}0} =
\sqrt{T_{\mathrm{e}0}/M_\mathrm{i}}$ is the reference plasma sound speed. 
Here, $M_\mathrm{i}$ denotes the ion mass and $T_{\mathrm{e}0}$ is 
the reference electron temperature.  The dependent variables are normalised by
$n_z \leftarrow n_z/n_{z0}$, $T_z \leftarrow T_z/T_{z0}$, $u_{z\parallel}
\leftarrow u_{z\parallel}/c_{\mathrm{s}0}$, $q_{z\parallel} \leftarrow
q_{z\parallel}/(n_{z0}T_{z0}c_{\mathrm{s}0})$, $\phi \leftarrow
(e\phi)/T_{\mathrm{e}0}$, $A_\parallel \leftarrow
A_\parallel/(\rho_{\mathrm{s}0}\beta_{\mathrm{e}0}B_0)$, where $n_{z0}$
represents the reference density, $T_{z0}$ is the reference temperature, $e$
denotes the elementary charge, $B_{0}$ represents the equilibrium magnetic flux
density, $\rho_{\mathrm{s}0}=c\sqrt{M_\mathrm{i}T_{\mathrm{e}0}}/(eB_0)$ is the
drift scale and $\beta_{\mathrm{e}0}=4\pi p_{\mathrm{e}0}/B_0^2$ is the reference
value for the electron dynamical beta. Here,
$p_{\mathrm{e}0}=n_{\mathrm{e}0}T_{\mathrm{e}0}$ denotes the reference electron
pressure.  The magnetic flux density is normalised by $B_0$. 

The main model parameters are the electron dynamical beta
$\beta_{\mathrm{e}0}$, the normalised drift scale
$\delta_0=\rho_{\mathrm{s}0}/a$ and the collisionality
$\nu_{\mathrm{e}0}=a/c_{\mathrm{s}0}\tau_{\mathrm{e}0}$, where
$\tau_{\mathrm{e}0}$ denotes the reference value for the Braginskii electron
collision time \cite{kendl10,scott05}.
The model dynamically evolves the fluctuating and the zonal 
and axisymmetric sideband components of the dependent variables.
Here, zonal denotes the flux-surface average and sideband denotes the
axisymmetric but non-zonal component.
The inner (source) radial boundaries for the axisymmetric part 
of the variables are given by zero Neumann conditions.
The inner radial boundaries for
the fluctuating part of the variables as well as the outer (sink) radial
boundaries are given by zero Dirichlet conditions.  The computational domain
includes an edge pedestal closed-flux-surface (CFS) region with consistent
quasi-periodic parallel-perpendicular boundary conditions and a
scrape-off-layer (SOL) region where the parallel boundary conditions represent a Debye
sheath limiter placed at the bottom side of a circular torus
\cite{ribeiro05,ribeiro08}.

The evolution of the profiles is self-consistently coupled to the magnetic
Shafranov equilibrium for circular flux surfaces.  Both the safety factor $q$
and the Shafranov shift are evolved in each time step \cite{scott06}.

The geometry is described in terms of field-aligned, unit-Jacobian Hamada
coordinates $(x,y_k,s)$ through
\begin{eqnarray}
   x &=& V = 2\pi^2R_0r^2, \\
   y_k &=& y-\alpha_k = q\theta-\zeta-\alpha_k, \\
   s &=& \theta \label{eq:ham}
\end{eqnarray}
where $V$ is the volume enclosed by the flux surface with major radius $R_0$
and minor radius $r$, and $\theta$ ($0\leq\theta<1$) and $\zeta$
($0\leq\zeta<1$) are the unit-cycle poloidal and toroidal Hamada angles (see
Ref.\ \cite{kendl10} for their definition). $V$ is given in units of $a^3$, and
$R_{0}$ and $r$ are normalised by $a$. In order to avoid magnetic shear
deformation of grid cells, the $y$-coordinate is shifted by
$\alpha_k=q\theta_k+\Delta\alpha_k$, i.e. $\Delta\alpha_k$ is chosen to make
$\bnabla x$ and $\bnabla y_k$ locally orthogonal at $\theta=\theta_k$
\cite{scott01}.

The initial magnetic equilibrium is computed from a prescribed safety factor
profile $q_0(x)$.  The temporal evolution of the Shafranov shift and $q(x)$ are
determined by the Pfirsch-Schl\"uter current and the associated axisymmetric 
component of $A_\parallel$ \cite{scott06}.
The Shafranov shift is incorporated into the
coordinate grid by modifying the metric elements according to the $s$-$\alpha$
model.  The resulting relevant part of the coordinate metric is given by
\begin{eqnarray}
   g^{xx} &=& \bnabla x \cdot \bnabla x
	  = (2\pi)^4\left(R_0r\right)^2 + \mathcal{O}(\varepsilon) 
   \label{eq:gxx} \\
   g^{yy}_k &=& \bnabla y_k \cdot \bnabla y_k 
            = \frac{q^2}{(2\pi r)^2} + \mathcal{O}(\varepsilon) 
   \label{eq:gyy} \\
   g^{xy}_k &=& \bnabla x \cdot \bnabla y_k 
            = 0 \quad \mathrm{at} \quad \theta=\theta_k 
   \label{eq:gxy}
\end{eqnarray}
where $g^{xx}$ and
$g^{yy}_{k}$ are given to lowest order in $\varepsilon=r/R_0$.

RMPs are included by adding a time-independent perturbation to the intrinsic parallel magnetic
potential (i.e. $A_{\parallel}\rightarrow A_{\parallel}+A_{\mathrm{p}}$). The
perturbation potential is defined as \cite{reiser05}
\begin{eqnarray}
    A_{\mathrm{p}} &=& - A_0 \sum\limits_{m}\left(-1\right)^{m}g(r)\cos[2\pi(m\theta-n\zeta)]\\
    &=& - A_0 \sum\limits_{m}\left(-1\right)^{m}g(r(x))\cos\{2\pi[ms-n(qs-y_k-\alpha_k)]\}
\end{eqnarray}
where $m$ and $n$ are the poloidal and toroidal mode number of the
perturbation, $A_0$ is an amplitude factor and $g(r)$ denotes the radial
envelope of the perturbation potential. 
For a perturbation including several poloidal mode numbers, the factor $(-1)^{m}$ can
be used to define the poloidal localization of the perturbation field. The
generation of RMPs by external perturbation coils requires RMPs to satisfy the
constraint of zero additional plasma current. Using Amp\`ere's law, this yields
to the Poisson equation
\begin{equation}
    -\nabla_{\perp}^{2}A_{\mathrm{p}}
    = -\frac{1}{J}\frac{\partial}{\partial x^{\mu}}Jg_{\perp}^{\mu\nu}\frac{\partial}{\partial x^{\nu}}A_{\mathrm{p}}
    = J_{\mathrm{p}}
    \equiv 0
    \label{eq:cfrmp}
\end{equation}
where $J$ is the coordinate Jacobian and $g^{\mu\nu}_{\perp}$ denotes metric
elements.  The time-dependent, global geometry implies that
Eq.\ (\ref{eq:cfrmp}) has to be solved numerically in each time step.

The initial state for the turbulence computations is based on typical ASDEX Upgrade (AUG)
L-mode profiles.  
The AUG plasma values are here used only to introduce a typical medium-size
tokamak reference scenario, and no direct comparison with experimental results
is at present intended on grounds of several idealisations present in the model,
like, for example, the use of simplified geometry and local (linearised) polarisation.

The mid-pedestal reference values for density, temperature
and magnetic equilibrium field are
$n_{\mathrm{e}0}=n_{\mathrm{i}0}=2\times10^{-19}\,\mathrm{m}^{-3}$,
$T_{\mathrm{e}0}=T_{\mathrm{i}0}=100\,\mathrm{eV}$ and $B_{0}=2.5\,\mathrm{T}$.
The resulting reference values for plasma beta, electron collisionality, drift
scale and Lundquist number are $\beta_{\mathrm{e}0}=6.4\times10^{-5}$,
$\nu_{\mathrm{e}0}=1.6\times10^{6}\,\mathrm{s^{-1}}$,
$\rho_{\mathrm{s}0}=5.8\times10^{-4}\,\mathrm{m}$ and $S_{0}=1.2\times10^{7}$.

We consider a circular toroidal tokamak with major radius
$R_{0}=1.65\,\mathrm{m}$ and minor radius $a=0.5\,\mathrm{m}$. The simulation
domain has a radial extension of $\Delta r=0.07\,\mathrm{m}$ around the
separatrix at $r=a$.  The initial safety factor profile is prescribed by
$q_{0}(r)=1.42+3.31(r/a)^{2}$ which yields $q$-values within the interval
$4.28\leq q_{0}\leq 5.21$ and a magnetic shear in the range
$1.34\leq\hat{s}_{0}\leq1.46$. The initial gradient lengths of density and temperature
are $L_{n}=0.07\,\mathrm{m}$ and $L_{T}=L_{\perp}=L_{n}/2$. 

We use a space resolution of $x \times y_{k} \times s = 64 \times 512 \times
16$ grid points, where $x$ denotes the radial, $y_{k}$ the perpendicular
(Clebsch angle) and $s$ the poloidal direction. 
The resulting resolution includes 
perpendicular scales down to twice the drift scale. The time resolution is
$0.002\,a/c_{\mathrm{s}}=1.4\times10^{-8}\,\mathrm{s}$. The initial pedestal
profiles for density and temperature are prescribed by
$n(r)/n_{0}=(L_{\perp}/L_{n})g_{0}(r)$ and
$T(r)/T_{0}=(L_{\perp}/L_{T})g_{0}(r)$ with $g_{0}(r)$ modelled as
$g_{0}(r)=0.5-0.5\sin\{2\pi[r-(a-\Delta r/4)]/\Delta r\}$ for $a-\Delta r/2\leq
r\leq a$.  The erosion of the pedestal profiles by radial transport is
counteracted by a time-independent source flux which models the radial inflow of
particles and heat from the core region.  Thus, the simulations saturate at a
statistically stationary turbulent state.

\begin{figure}
	\subfloat[$A_{0}=100$]{\includegraphics{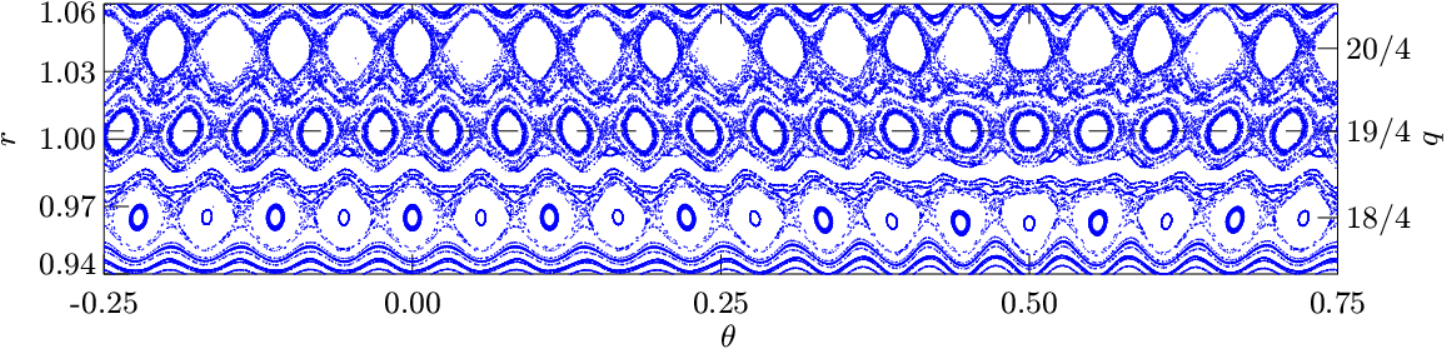}}\\
	\subfloat[$A_{0}=200$]{\includegraphics{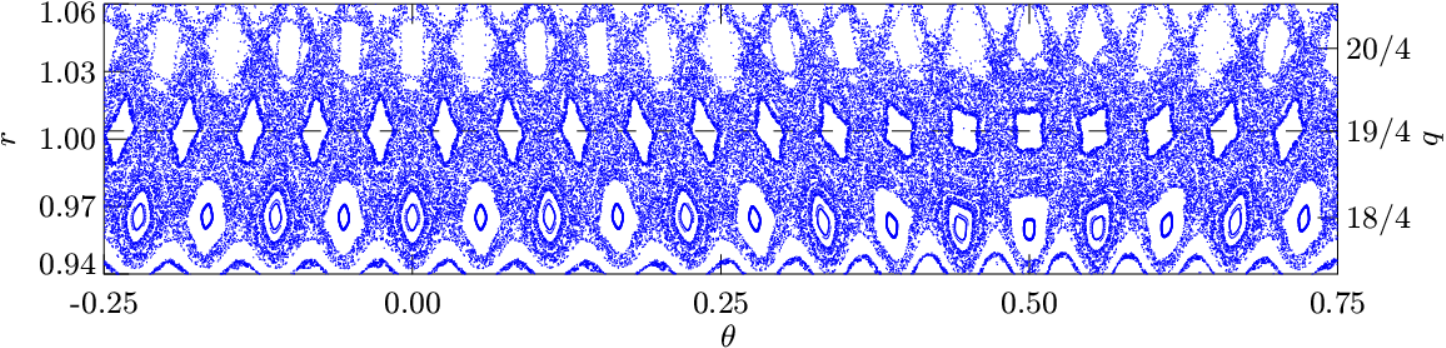}}\\
	\subfloat[$A_{0}=400$]{\includegraphics{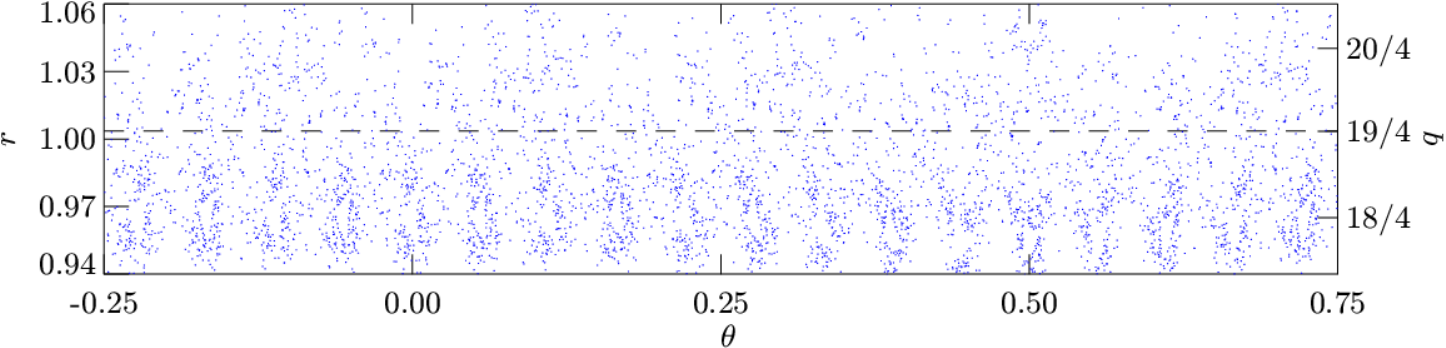}}
    \caption{\sl
	Poloidal Poincar\'e sections of the vacuum RMP fields for the
	perturbation amplitudes ($A_{0}=100$, $200$, $400$) used in the
        simulations. 40 magnetic field 
	lines were followed over 4000 toroidal turns. In (c), most of the field
	lines leave the simulation domain after some 100 toroidal turns. 
        The dashed lines mark the $q=19/4$ position close to the separatrix (at $r=1$).
    }
    \label{fig:pcv}
\end{figure}

The computations include RMP fields with three different perturbation
amplitudes $A_{0}$ and helicity components $m/n=18/4$, $m/n=19/4$ and
$m/n=20/4$. The magnetic perturbation potential is prescribed to be equal on
both radial boundaries, with $g(a-\Delta r/2)=g(a+\Delta r/2)=1$. The applied
perturbation amplitudes correspond to $9.3\times10^{-6}\,\mathrm{Tm}$
($A_{0}=100$), $1.9\times10^{-5}\,\mathrm{Tm}$ ($A_{0}=200$) and
$3.7\times10^{-5}\,\mathrm{Tm}$ ($A_{0}=400$) and yield magnetic perturbation
fields of order $10^{-3}\,\mathrm{T}$.  The RMP fields are located on the
high-field side at $\theta=0.5$. Fig.~\ref{fig:pcv} shows poloidal Poincar\'e
sections of the vacuum RMP fields.  For $A_{0}=100$ the RMP-induced islands at
resonant flux surfaces are well separated.  For $A_{0}=200$ islands overlap and
ergodic regions between the islands are formed.  A further increase of the
perturbation amplitude to $A_{0}=400$ results in a strongly ergodised magnetic
field.

\section{L-mode turbulence simulations}

The simulations are evaluated in terms of space and time averages over
statistically stationary turbulent states. Time averages are taken over an
interval of $\Delta t=1000\,a/c_{\mathrm{s}}=7.2\times10^{-3}\,\mathrm{s}$.
For the evaluation the dependent variables are separated as 
\begin{equation}
    f(x,y_{k},s,t)=\langle f(x,y_{k},s,t)\rangle_{t}+\tilde{f}_t(x,y_{k},s,t)
    \label{eq:ftf}
\end{equation}
where $\langle f(x,y_{k},s,t)\rangle_{t}$ and $\tilde{f}_t(x,y_{k},s,t)$ denote
the stationary and the temporally fluctuating part of a
dependent variable $f(x,y_{k},s,t)$. Spatial fluctuations of a dependent
variable are computed with respect to its toroidal mean as
$\tilde{f}_x(x,y_{k},s,t)=f(x,y_{k},s,t)-\langle f(x,y_{k},s,t)\rangle_{y_{k}}$.

\subsection{Screening of RMPs by plasma response currents}

\begin{figure}[h]
    \begin{center}
	\subfloat[$A_{0}=100$]{\includegraphics{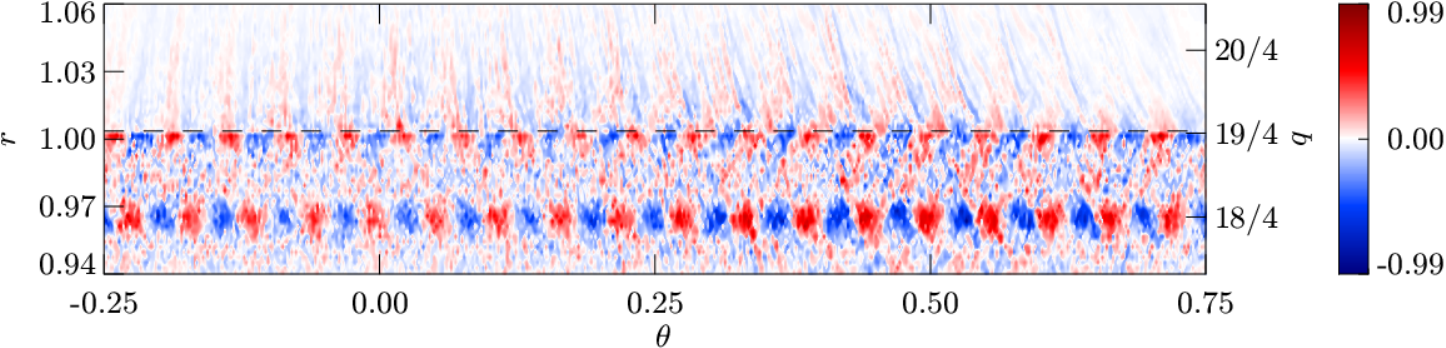}}\\
	\subfloat[$A_{0}=200$]{\includegraphics{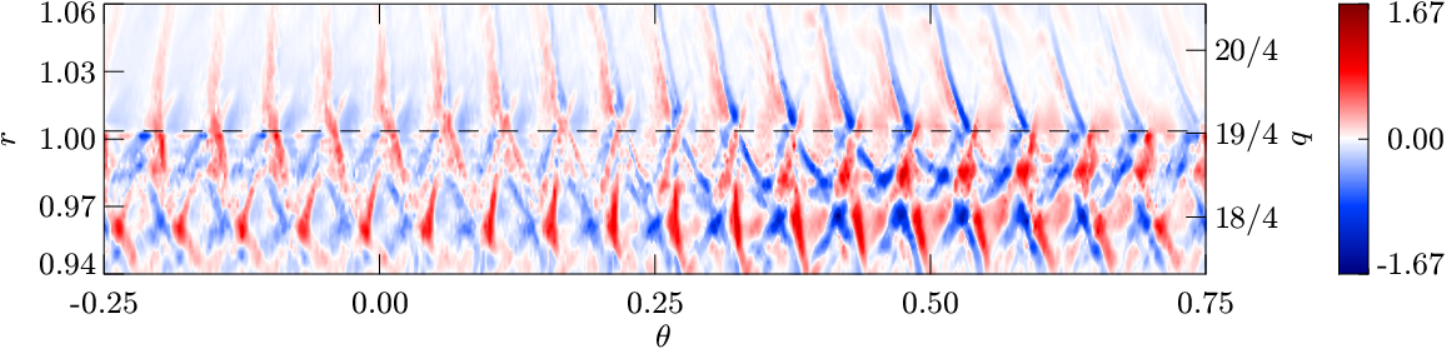}}\\
	\subfloat[$A_{0}=400$]{\includegraphics{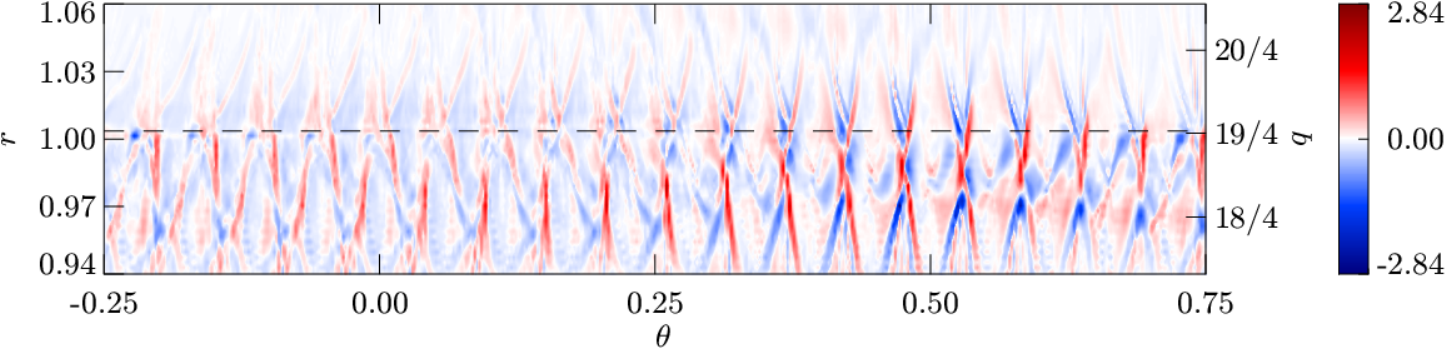}}
    \end{center}
    \caption{\sl 
	RMP-induced variation of the stationary part of the parallel current
	fluctuations in the poloidal plane. The fluctuations were computed by
	subtracting the toroidal mean as
	$\tilde{J}_{\parallel}=J_{\parallel}-\langle
	J_{\parallel}\rangle_{y_{k}}$. 
        The dashed lines mark the $q=19/4$ position.
    }
    \label{fig:jpl}
\end{figure}

The plasma response to externally applied RMPs consists of parallel currents
which alter the vacuum structure of the perturbation fields.
Fig.~\ref{fig:jpl} shows the RMP-induced variation in the stationary component
of the parallel plasma current fluctuations in the poloidal plane. It is found
that RMPs give rise to current fluctuations at resonant flux surfaces. The
fluctuation amplitudes increase quasi-linearly with the perturbation amplitude.
Due to lower fluctuation amplitudes in the SOL, the effect is more pronounced
in the CFS region.

\begin{figure}
    \begin{center}
	\subfloat[$A_{0}=100$]{\includegraphics{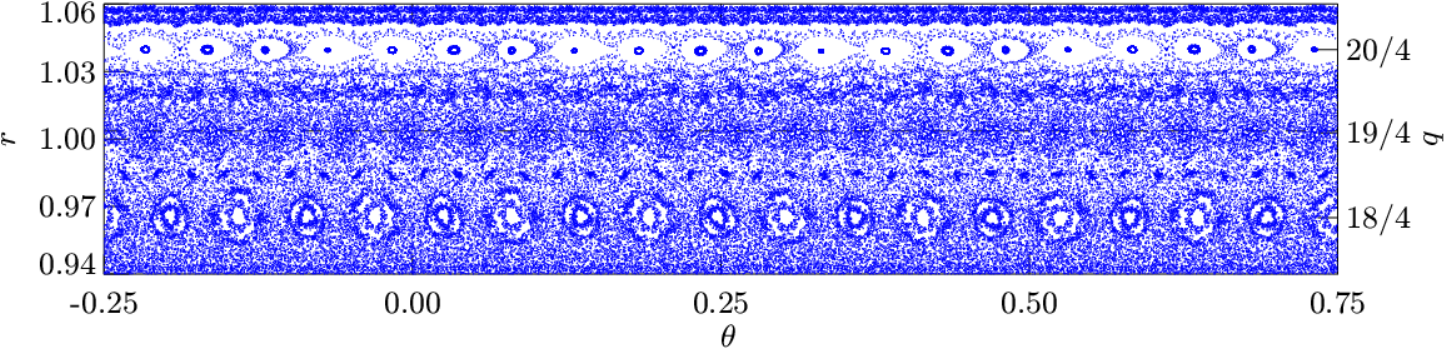}}\\
	\subfloat[$A_{0}=200$]{\includegraphics{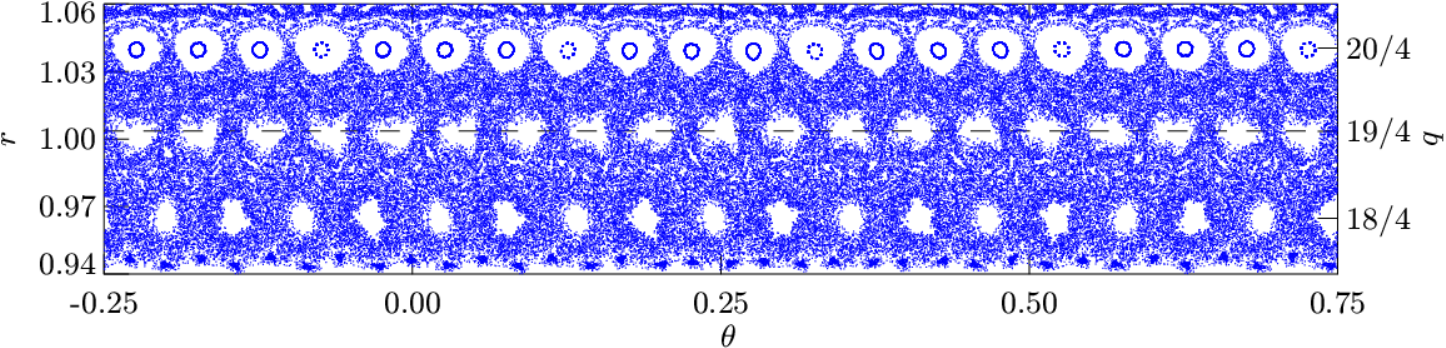}}\\
	\subfloat[$A_{0}=400$]{\includegraphics{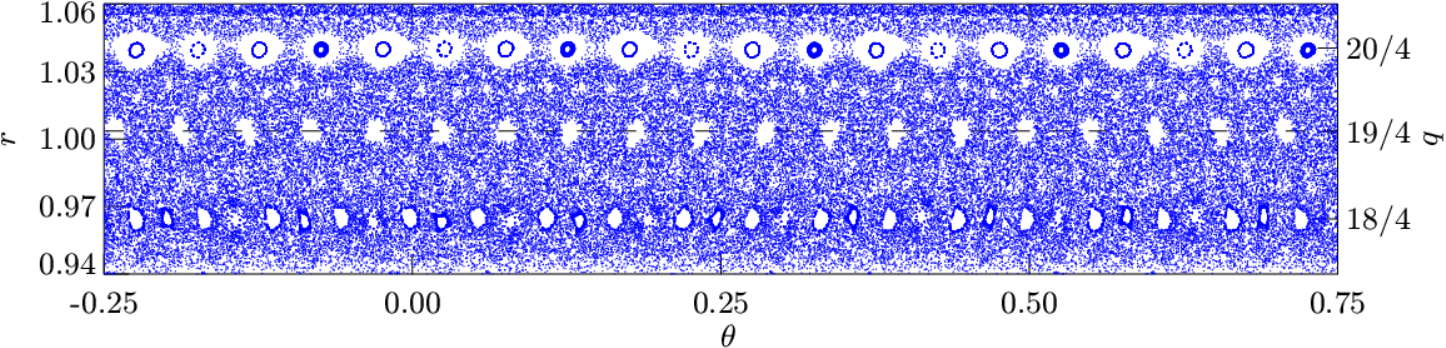}}
    \end{center}
    \caption{\sl
	Poloidal Poincar\'e sections of the stationary part of the magnetic
	flutter field for various RMP amplitudes. 40 magnetic field
	lines were followed over 4000 toroidal turns.  
        The dashed lines mark the $q=19/4$ position.
    }
    \label{fig:pci}
\end{figure}

The corresponding Poincar\'e plots, computed from the stationary component of
the magnetic flutter field, are shown in Fig.\ \ref{fig:pci}.  The magnetic
island structure exhibits the mode number of the imposed RMP fields but the
islands are poloidally shifted by about half a poloidal island. Moreover, the
radial extension of the islands is reduced. Thus, the simulations show that the
plasma response currents are very effective at screening the vacuum RMP fields.
The magnetic islands associated with the vacuum RMP fields are closed and
reopened with decreased amplitude at a poloidally shifted position. The
screening effect can furthermore be quantified by the amplitude of the
perpendicular magnetic flutter. For $A_{0}=400$ the magnetic flutter is reduced
by 73\,\% with respect to the vacuum RMP fields.

\subsection{Thermal state variables}

\begin{figure}
    \begin{center}
	\subfloat[]{\includegraphics{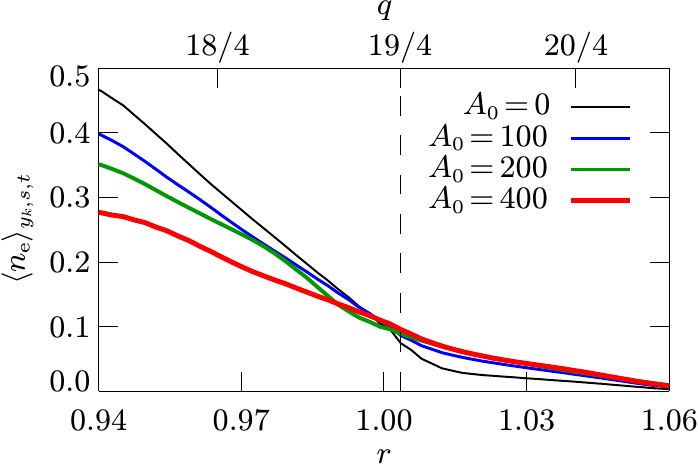}}\hfill
	\subfloat[]{\includegraphics{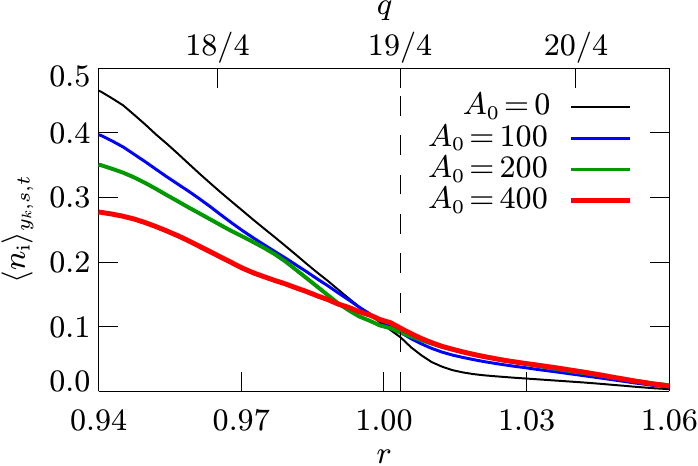}}\\
	\subfloat[]{\includegraphics{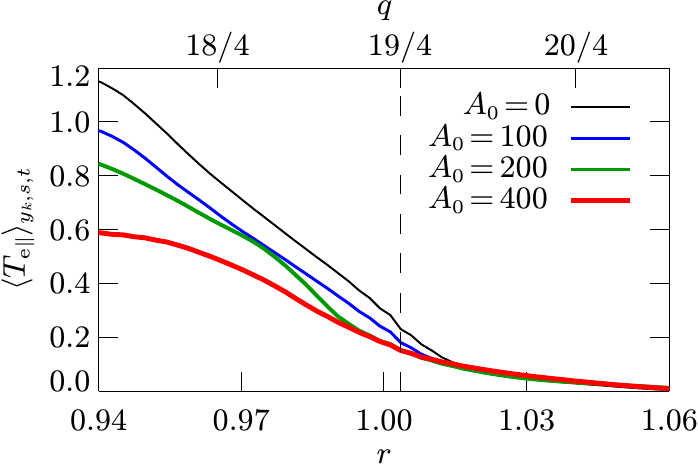}}\hfill
	\subfloat[]{\includegraphics{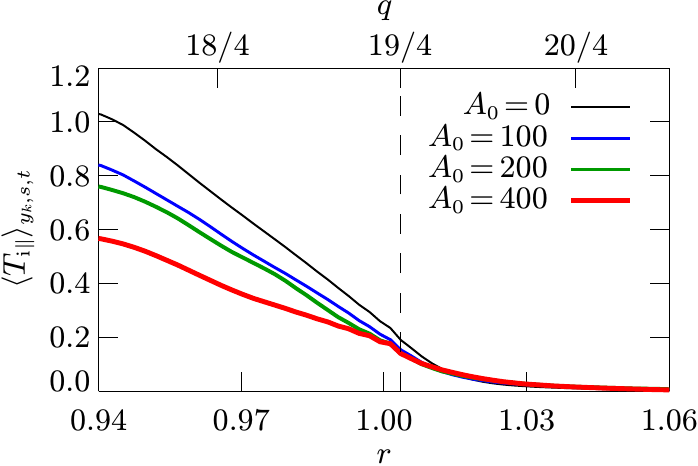}}\\
	\subfloat[]{\includegraphics{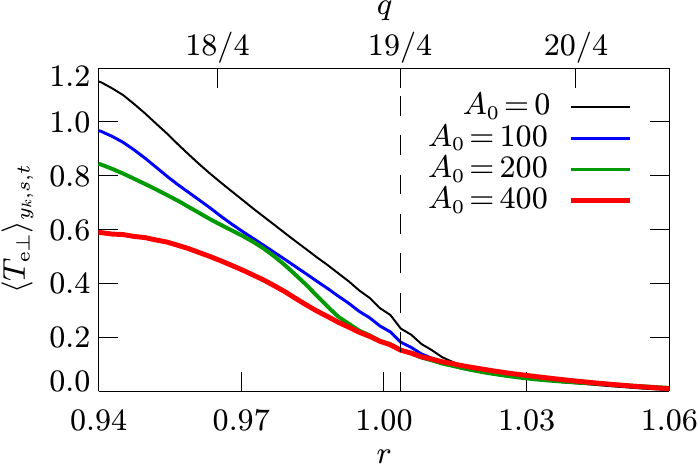}}\hfill
	\subfloat[]{\includegraphics{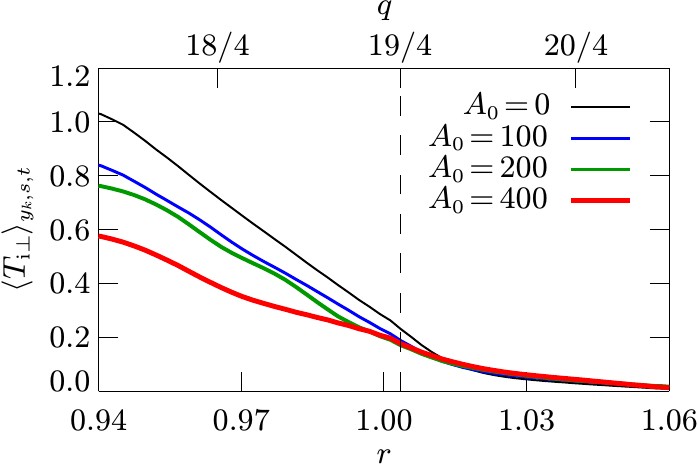}}
    \end{center}
    \caption{\sl
	Time- and flux-surface-averaged radial profiles of (a) electron
	density, (b) ion density, (c) parallel electron temperature, (d)
	parallel ion temperature, (e) perpendicular electron temperature and
	(f) perpendicular ion temperature for various RMP amplitudes.
        The dashed lines mark the $q=19/4$ position.
    }
    \label{fig:tsv}
\end{figure}

In the following, we discuss the effects of RMPs on thermal state variables
($n_{\mathrm{e}}$, $n_{\mathrm{i}}$, $T_{\mathrm{e}\parallel}$,
$T_{\mathrm{e}\perp}$, $T_{\mathrm{i}\parallel}$, $T_{\mathrm{i}\perp}$).
Fig.~\ref{fig:tsv} shows time-averaged zonal profiles of
densities and temperatures. For each of the thermal state variables, RMPs give
rise to a flattening of the profiles in the CFS-region. In the SOL, the density
profiles are slightly increased by RMPs, while the temperature profiles remain
nearly unchanged. Note that the changes in the profiles increase with
increasing RMP-amplitude. Furthermore, the changes are not confined to resonant
flux surfaces but involve the entire radial simulation domain.

\begin{figure}
    \begin{center}
	\subfloat[Total density, CFS]{\includegraphics{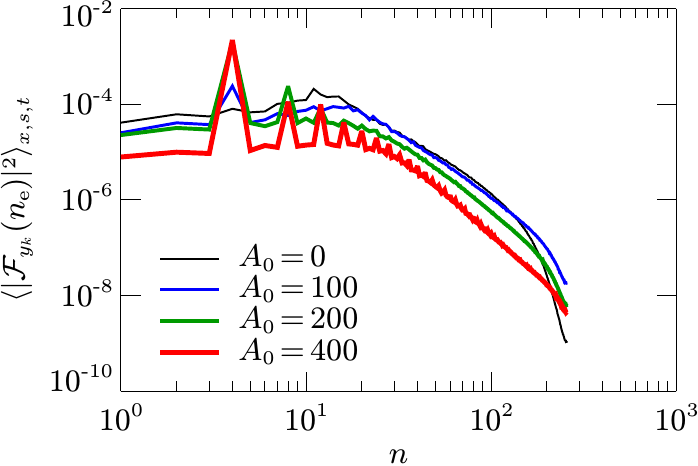}}\hfill
	\subfloat[Total density, SOL]{\includegraphics{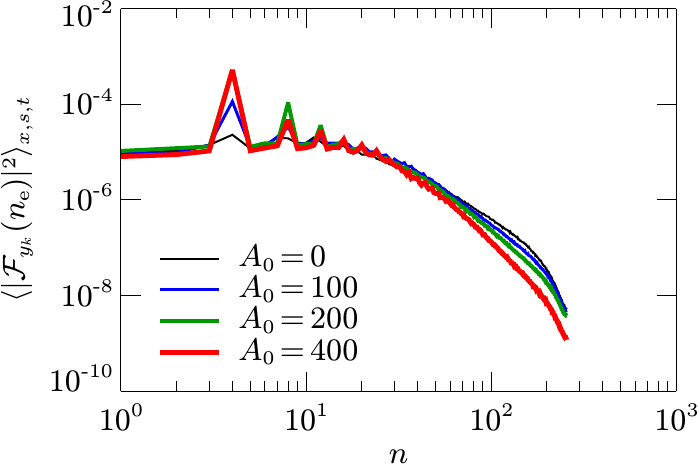}}\\
	\subfloat[Stationary part, CFS]{\includegraphics{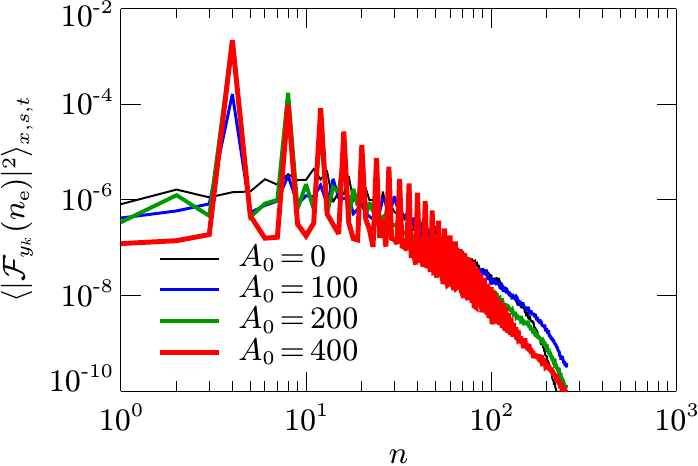}}\hfill
	\subfloat[Stationary part, SOL]{\includegraphics{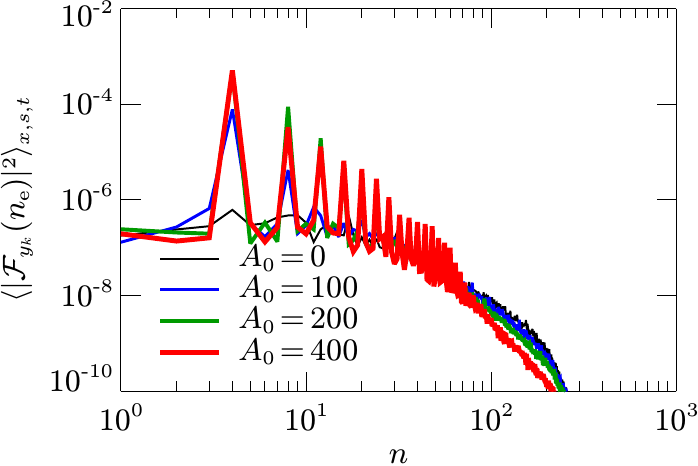}}\\
	\subfloat[Temporally fluctuating part, CFS]{\includegraphics{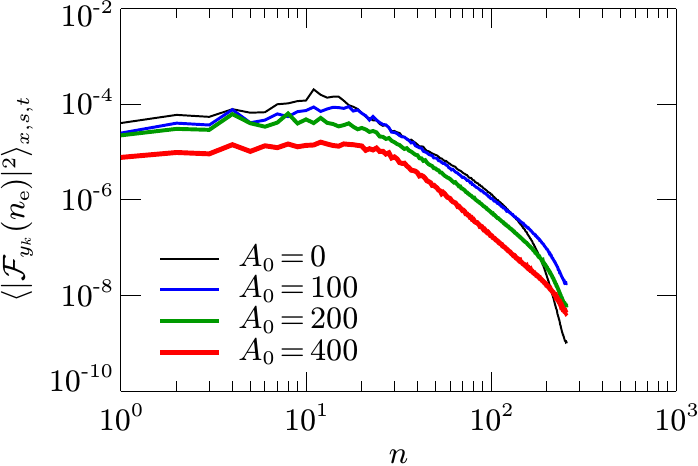}}\hfill
	\subfloat[Temporally fluctuating part, SOL]{\includegraphics{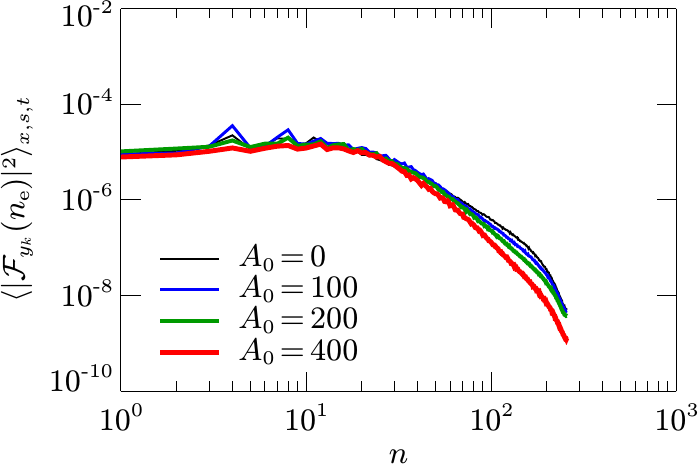}}
    \end{center}
    \caption{\sl
	Time- and space-averaged toroidal mode number spectra of the electron
	density for various RMP amplitudes. The (a, b) total density
	and the (c, d) stationary and (e, f) temporally fluctuating parts in
	both CFS region an SOL are shown.
    }
    \label{fig:nsp}
\end{figure}

Fig.~\ref{fig:nsp} illustrates RMP-induced changes in toroidal mode number
spectra of the electron density.  The total density as well as the stationary
and the temporally fluctuating parts according to Eq.\ (\ref{eq:ftf}) are shown
for both CFS region and SOL.  The spectra of the total fluctuations in the CFS
region indicate that RMPs give rise to the formation of density structures with
mode numbers which are resonant with the applied RMP fields. While these
resonant structures are especially pronounced in the stationary part, they do not
occur in the temporally fluctuating part. 

Thus, the imposed resonant structures are static and this is reflected in the
resulting stationary component, but not in the fluctuations \cite{reiser05}.  
The temporally fluctuating part of the density in the CFS region decreases
with increasing perturbation amplitude. 
Accordingly, the RMP fields amplify the resonant modes but attenuate other
modes. The SOL exhibits the same stationary resonant structures, although at lower
amplitudes. In contrast to the CFS region, the temporally fluctuating parts of
the density are not significantly attenuated in the SOL.  The effects of RMPs
on other thermal state variables are found to be very similar. 

The results on the static component are in agreement with earlier work by
Reiser {\sl et al.} on RMP effects on local isothermal drift-Alfv\'en edge
turbulence, who first found that ``strong resonant effects can be attributed
to (quasi-) static contributions of the perturbations'' \cite{reiser05}. 


The fluctuations in the thermal state variables reflect the three-dimensional
magnetic equilibrium imposed by RMPs. The RMP-induced toroidal variation of
the magnetic equilibrium modifies the diamagnetic equilibrium current
$\vec{J}_{*}\propto\vec{B}\times\bnabla p$.  This impacts the electric
potential and consequently the thermal state variables \cite{reiser05}.

\subsection{Convective and magnetic transport}

In order to address the question of how RMPs influence the radial transport, we
compare the RMP-induced effects on the convective and magnetic transport of
electron heat, in differential form given by
\begin{eqnarray}
    \mathrm{d} Q_{\mathrm{E},\mathrm{e}} &=&
    {3 \over 2} p_{\mathrm{e}}u_{\mathrm{E}}^{x}\mathrm{d}y_{k}\mathrm{d}s \label{eq:qee} \\
    \mathrm{d} Q_{\mathrm{M},\mathrm{e}} &=&
    q_{\mathrm{e}\parallel}b^{x}\mathrm{d}y_{k}\mathrm{d}s \label{eq:qem}
\end{eqnarray}
where $u_E^x = B^{-2} \vec{B} \times \vec{\nabla} \phi \cdot \vec{\nabla} x$
denotes the radial component of the $E\times B$ drift velocity 
and $b^x = - B^{-2} \vec{B} \times \vec{\nabla} A_{||} \cdot \vec{\nabla} x$.
The radial transport of ion heat and density fluctuations are
found to exhibit similar characteristics.

\begin{figure}
    \begin{center}
	\subfloat[]{\includegraphics{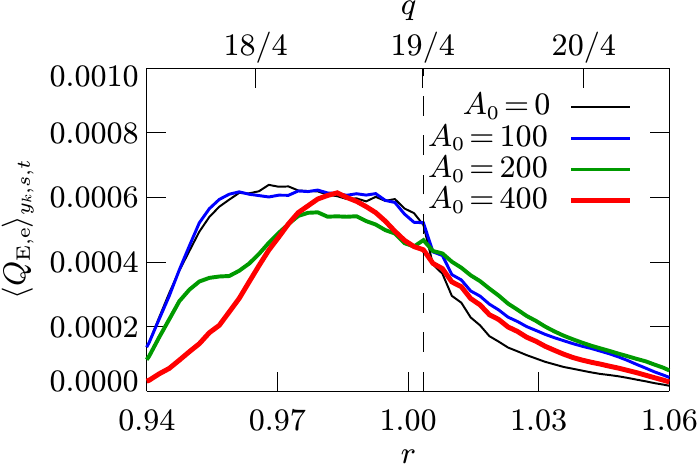}}\hfill
	\subfloat[]{\includegraphics{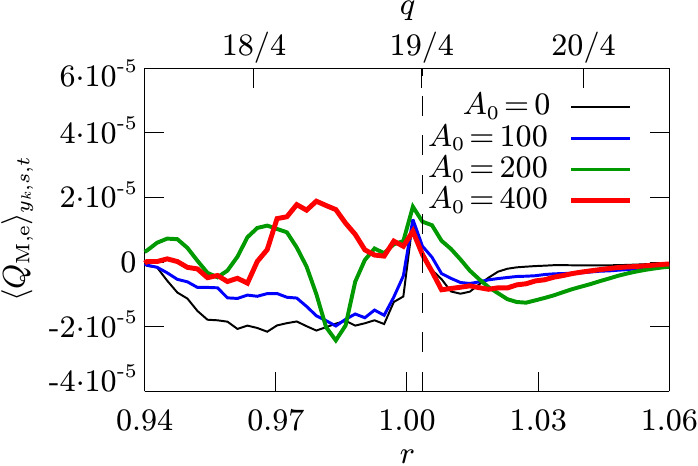}}
    \end{center}
    \caption{\sl
	Time- and flux-surface-averaged radial profiles of the (a) convective
	and (b) magnetic radial transport of electron heat for various
	RMP amplitudes. 
    }
    \label{fig:qem}
\end{figure}

Fig.~\ref{fig:qem} shows time- and flux-surface-averaged radial profiles of the
convective and magnetic radial transport of electron heat for various
RMP amplitudes.  A comparison of the magnitudes shows that the magnetic
transport is negligibly small for all perturbation amplitudes. Even for ergodic
vacuum perturbation fields, the radial transport is dominated by turbulent
convection. Note that this is in agreement with the observed screening of the
vacuum perturbation fields \cite{reiser09}.

Considering Fig.\ \ref{fig:qem} we can conclude that the sum of convective and
magnetic transport varies somewhat with the applied perturbation amplitude.  
As all simulations include the same source flux and the physical dissipation
terms are found to be nearly unchanged across different simulations, this
effect can be ascribed to the varying transport losses through the outer edge
boundary layer. 

\begin{figure}
    \begin{center}
	\subfloat[Total transport, CFS]{\includegraphics{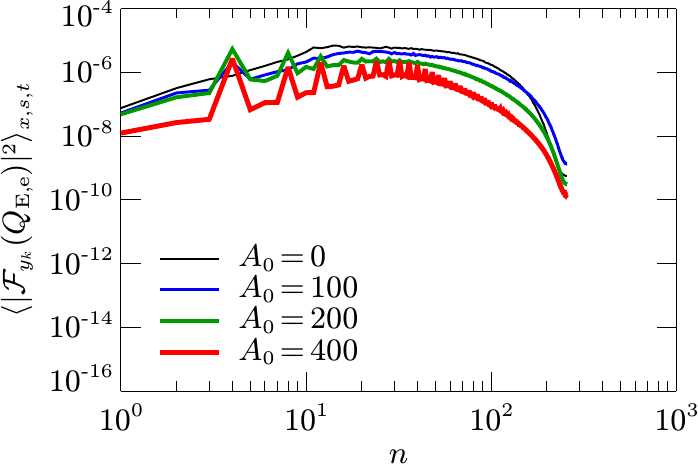}}\hfill
	\subfloat[Total transport, SOL]{\includegraphics{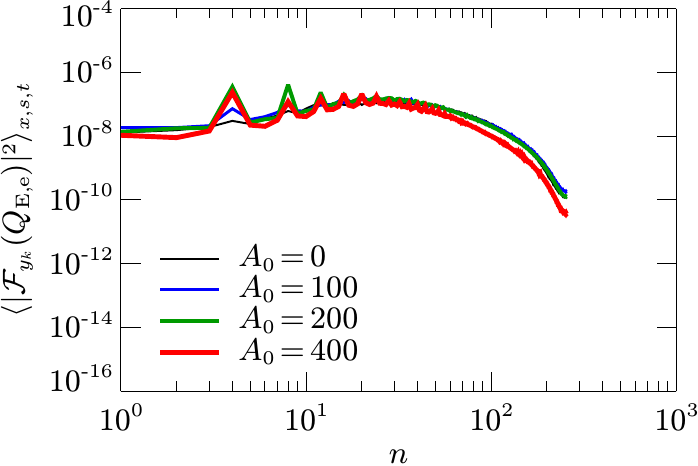}}\\
	\subfloat[Stationary part, CFS]{\includegraphics{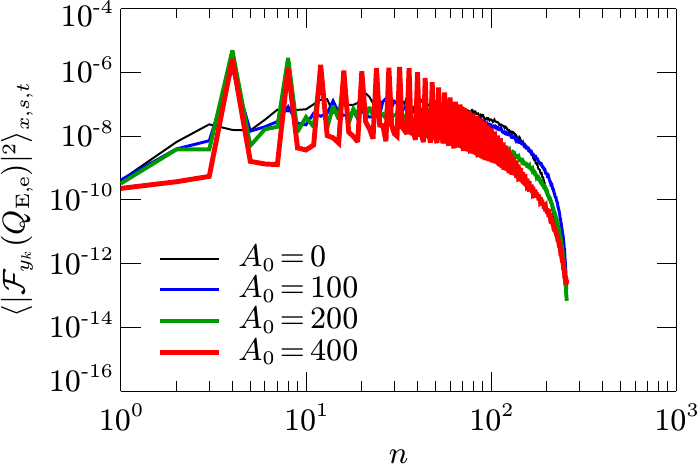}}\hfill
	\subfloat[Stationary part, SOL]{\includegraphics{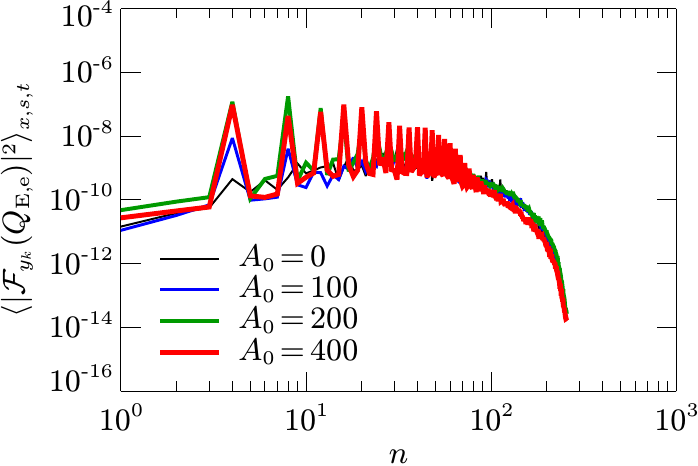}}\\
	\subfloat[Temporally fluctuating part, CFS]{\includegraphics{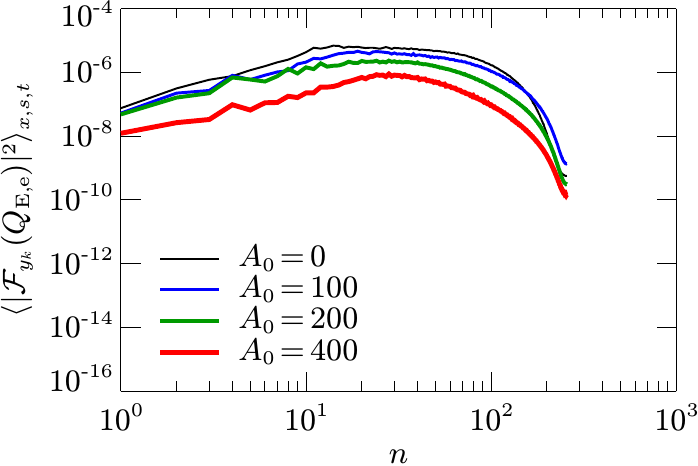}}\hfill
	\subfloat[Temporally fluctuating part, SOL]{\includegraphics{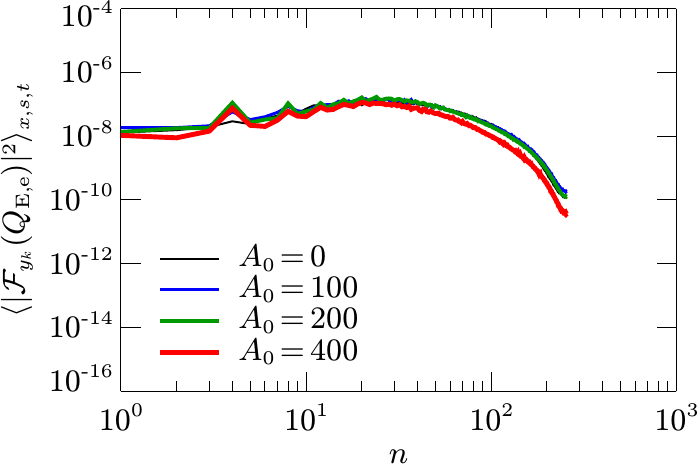}}
    \end{center}
    \caption{\sl
	Time- and space-averaged toroidal mode number spectra of the convective
	transport of electron heat for various RMP amplitudes. The (a,
	b) total transport and the (c, d) stationary and (e, f) temporally
	fluctuating parts in both CFS region an SOL are shown.
    }
    \label{fig:qsp}
\end{figure}

In the following we consider the modification of the convective transport
by RMPs.  Fig.~\ref{fig:qsp} illustrates the toroidal structure of the time-
and space-averaged convective transport of electron heat. Toroidal mode number
spectra of the total transport ($p_{\mathrm{e}}\vec{u}_{\mathrm{E}}$) as well
as the contributions resulting from the stationary ($\langle
p_{\mathrm{e}}\rangle_{t}\langle\vec{u}_{\mathrm{E}}\rangle_{t}$) and
temporally fluctuating ($\tilde{p}_{\mathrm{e}}\tilde{\vec{u}}_{\mathrm{E}}$) parts
 of the dependent
variables are shown for both CFS region and SOL.  In the CFS region, the
toroidal transport is composed of the smooth temporally fluctuating part and
the comb-like structure resulting from the stationary contribution. The temporally
fluctuating part decreases with increasing perturbation amplitude. This
decrease is partly compensated by the stationary part.  Hence, the application of
RMPs gives rise to a reorganisation of the total radial transport in terms of
stationary convection cells. In the SOL, the effects of RMPs are similar.
However, the RMP-induced reduction of the turbulent transport is less
pronounced in the SOL.

\subsection{Poloidal plasma rotation and GAMs}

\begin{figure}[h]
    \begin{center}
	\includegraphics{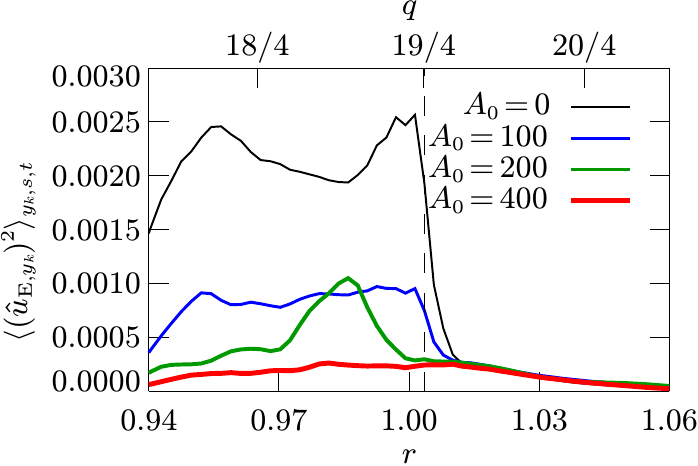}
    \end{center}
    \caption{
	Time- and flux-surface-averaged radial profiles of the squared poloidal
	$E\times B$ rotation velocity for various RMP amplitudes. 
        The dashed lines mark the $q=19/4$ position.
    }
    \label{fig:vey}
\end{figure}

RMPs are found to influence the radial profiles of the electric potential. In
the following, we quantify the associated changes in the poloidal $E\times B$
rotation velocity
$\hat{u}_{\mathrm{E}}^{y_{k}}=\vec{u}_{\mathrm{E}}\cdot\hat{\vec{\mathrm{e}}}_{y_{k}}$,
where $\hat{\vec{\mathrm{e}}}_{y_{k}}$ denotes the unit vector perpendicular to
both $\vec{\mathrm{e}}_{s}$ and $\vec{\mathrm{e}}_{x}$.  Fig.~\ref{fig:vey}
shows time- and flux-surface averaged radial profiles of the squared poloidal
$E\times B$ velocity.  It is shown that the application of RMPs involves a
reduction of the poloidal plasma rotation.  Note that the average poloidal
velocity in the SOL is small so that the changes mainly concern the CFS region.  
The presented simulation setup does not allow to draw clear conclusions about the
effects of RMPs on geodesic acoustic modes (GAMs).

\subsection{Summary and discussion of turbulence results}

The effects of non-axisymmetric RMPs were investigated by numerical simulations
employing the nonlinear gyrofluid electromagnetic model GEMR.  The simulation
setup was arranged for typical L-mode conditions of the AUG tokamak. RMPs were
implemented in terms of a magnetic perturbation potential satisfying the
constraint of zero additional plasma current. The simulations were performed
for three amplitudes of a multiple RMP field including the helicity components
$m/n=18/4$, $m/n=19/4$, and $m/n=20/4$. The simulation results were discussed
in terms of time  and space averaged quantities. For the evaluation, the
dependent variables were separated into an RMP-induced stationary component and a
temporally fluctuating, turbulent contribution.

The main findings can be summarised as follows:
\begin{enumerate}
    \item RMPs give rise to the formation of resonant, parallel plasma response
	currents which are radially-localised around resonant magnetic flux
	surfaces \cite{reiser09}.
    \item The vacuum RMP fields are effectively screened by the plasma response.
	The magnetic islands associated with the vacuum RMP fields are closed
	and replaced by islands which are poloidally shifted by half a poloidal
	island width.
    \item The magnitude of the intrinsic magnetic flutter is reduced by up to
	73\,\% with respect to the vacuum RMP fields.  Radial and poloidal
	profiles of the intrinsic magnetic flutter significantly differ from
	the imposed vacuum RMP fields.
    \item RMPs cause a non-localised flattening of density and temperature profiles.
    \item RMPs reduce the temporally fluctuating part of the electron density.
	This decrease is compensated by an RMP-induced formation
	of stationary, resonant density perturbations which reflect the
	three-dimensional magnetic equilibrium imposed by RMPs. 
    \item RMPs give rise to the formation of stationary convection cells which
	compensate the RMP-induced decrease of the transport associated with
	the temporally fluctuating parts of the dependent variables.
    \item The magnetic transport is negligibly small compared to the convective
	transport. Ergodic vacuum perturbation fields do not give rise to an
	increased magnetic transport. This is in agreement with the 
	screening of RMPs by the plasma response.
    \item RMPs involve an attenuation of the poloidal $E\times B$ rotation.
\end{enumerate}

The simulations were repeated for a multiple helicity RMP field located on the
low-field side at $\theta=0$, and a single RMP exhibiting no poloidal localisation.  
We found that the main effects (formation of resonant plasma fluctuations,
reorganisation of the turbulent transport in terms of resonant convection
cells, negligible magnetic transport) do not depend on either the poloidal
localisation or the number of included helicity components of the applied RMP
field.

In the present GEMR modeling, the major novel results concern the
self-consistent temperature dynamics of both electrons and ions. Thus,
convective and magnetic transport can be evaluated in a self-consistent model.

The screening of RMPs by plasma response currents was recently discussed in
Refs.\ \cite{becoulet12,reiser09}. In both works, the screening effect was
ascribed to the formation of radially-localised current layers in phase with
the RMP field. Moreover, out-of-phase currents leading to the growth of
poloidally shifted magnetic islands were reported \cite{reiser09}.  Our
simulations reproduce both the complete screening of the vacuum perturbation
fields by in-phase currents and the formation of poloidally shifted magnetic
islands by out-of-phase currents. 
The RMP-induced formation of stationary density fluctuations and resonant
convection cells was previously reported in
Refs.\ \cite{beyer02,reiser05,reiser07} and is in good agreement with our
results. Furthermore, our results confirm an RMP-induced reduction of the
zonal averaged poloidal $E\times B$ velocity as reported in
Ref.\ \cite{beyer02}.  By contrast, our simulation setup does not lead to an RMP-induced
increase of GAMs as found in \cite{reiser08}. Note however that the results in
\cite{reiser08} 
were obtained by particular adapted equations.

The RMP-induced decrease of the zonal density and temperature gradients in our
simulations could be an important indication on the mechanism governing the
mitigation of ELMs by RMPs. 
Edge turbulence models with self-consistent profiles do not exhibit transport
barriers when no further mechanisms are present. 
Unlike MHD interchange models, the two-fluid and gyrofluid models specifically
contain finite-beta drift-wave and ion temperature effects. 
These nonlinear drift wave instabilities always supersede linear
instabilities, hence there is no threshold character \cite{scott05b}. 
Thus, the effects of RMPs on H-mode configurations can not be studied in a
self-consistent manner. Assuming that RMP-induced changes are similar in L- and
H-mode plasmas, the decrease of the zonal pressure gradient could explain the
mitigation and suppression of pressure-gradient-driven ELMs.

In sum, we can conclude that our simulation results generally agree with
previous studies. Significant discrepancies can be ascribed to differences in
the basic model assumptions (local flux tube model without SOL versus global
geometry model including profile evolution and SOL). 
The heat transport associated with a self-consistent temperature dynamics (not
included by previous models) was found to exhibit similar characteristics to
the density transport. Moreover, the magnetic contribution to the radial
transport was found to be negligibly small even for strongly ergodised vacuum
RMP fields.

\section{Ballooning mode burst simulations}

In the following we analyse the use of computational setups within the present
gyrofluid approach to model ELMy H-mode like scenarios including RMP fields.

The computations discussed in this section are based on the ideal ballooning
unstable simulation setup described in Refs.\ \cite{kendl10,peer13}. 
The initial safety factor profile is defined as $q_{0}=1.41+3.29\,(r/a)^{2}$,
so that the flux surface with $q=m/n=18/4$ coincides with the position of the
steepest gradients at $r/a=0.97$.  

The local parameters, taken as mid pedestal values, correspond to electron and
ion temperatures $T_e = 300$~eV, $T_i = 360$~eV, densities $n_e = n_i = 2.5
\cdot 10^{19}$~m$^{-3}$, magnetic field strength $B=2.0$~T, major torus radius
$R = 1.65$~m, aspect ratio $R/a=3.3$, perpendicular temperature gradient
length $L_T = L_{\perp}=3.0$~cm, density gradient length $L_n = 6.0$~cm.
The radial domain of the simulations covers a range of $L_{\perp}$
on either side of the LCFS. 
As above, $n_y=512$ perpendicular mesh points are used, and the radial domain
with $n_x=64$ spans the plasma edge region between the H-mode pedestal top,
with plasma core parameters as inner boundary values, and the 
outer bounded scrape-off layer region ($r/a=1\pm 0.06$). 
This translates to a ratio $\delta = \rho_s / a = 0.0026$ between ion gyroradius 
$\rho_s$ and minor torus radius $a$, and a local plasma beta of $\beta_{e0} = 7.5
\cdot 10^{-4}$.

The initial profiles now correspond to a typical AUG (unmitigated)
ELMy H-mode scenario. We apply the RMP field (as described at the end of
sec.~2) to the initialised profiles.  

The RMP amplitudes are gives the values
$1.0 \cdot 10^{-7} \, \mathrm{Tm}$ ($A_{0}=0.1$), 
$1.0 \cdot 10^{-6} \, \mathrm{Tm}$ ($A_{0}=1$), 
$1.0 \cdot 10^{-5} \, \mathrm{Tm}$ ($A_{0}=10$), 
$2.0 \cdot 10^{-5} \, \mathrm{Tm}$ ($A_{0}=20$), and
$4.0 \cdot 10^{-5} \, \mathrm{Tm}$ ($A_{0}=40$).
The corresponding magnetic perturbation fields are of order
$10^{-6} \, \mathrm{T}$ ($A_{0}=0.1$), 
$10^{-5} \, \mathrm{T}$ ($A_{0}=1$), and
$10^{-4} \, \mathrm{T}$ ($A_{0}=10$-$40$), 

In the absence of RMP fields the initialised pedestal profiles are ideal ballooning
unstable and show single ELM-like bursts as discussed in Refs.\ \cite{kendl10,peer13}. 
In the following, effects of added RMP fields on this ideal ballooning blow-out scenario
are studied.

\subsection{Effects of RMPs on ideal ballooning unstable H-mode states}

We consider the ``H-mode ELM'' scenario of Refs.\ \cite{kendl10,peer13}
and focus on the evolution of an ideal ballooning blow-out, not including
any profile sustaining sources. In the RMP-free case, the prepared
pedestal state evolves into an ideal ballooning instability with toroidal mode
number $n=7$. If RMPs are applied, the plasma adjusts to the RMP-induced
three-dimensional magnetic equilibrium \cite{reiser05}, and the most unstable
mode number associated with the RMP-free nominal case ($n=7$) is found to be
replaced by a mode number which is resonant with the perturbation fields.
Below, we refer to this ideal ballooning scenarios as an ELM model, while
recognising its provisional status.

\begin{figure}
    \begin{center}
	\subfloat[$A_{0}=0$  \label{fig:specnfa}]{\includegraphics[width=71mm]{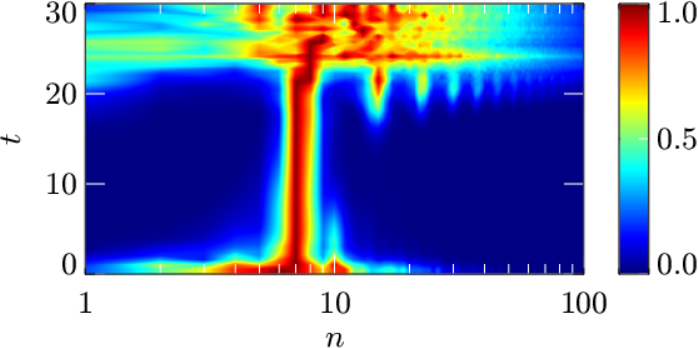}}\hfill
	\subfloat[$A_{0}=0.1$\label{fig:specnfb}]{\includegraphics[width=71mm]{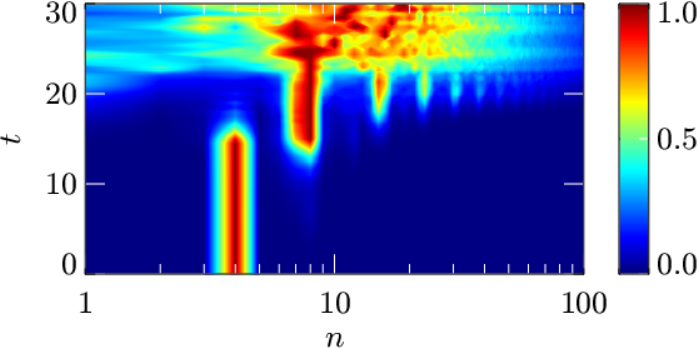}}\\
	\subfloat[$A_{0}=1$  \label{fig:specnfc}]{\includegraphics[width=71mm]{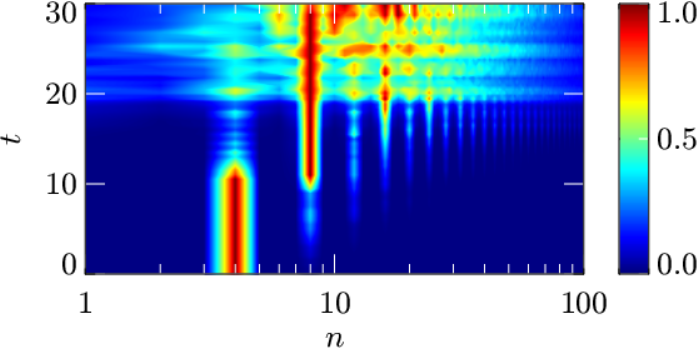}}\hfill
	\subfloat[$A_{0}=10$ \label{fig:specnfd}]{\includegraphics[width=71mm]{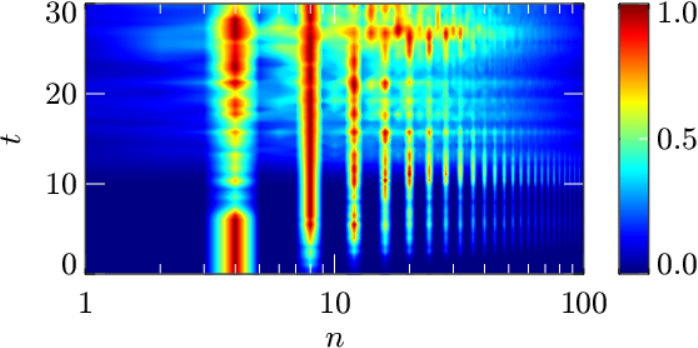}}
    \end{center}
    \caption{\sl
	Space-averaged toroidal mode number spectra associated with the time
	evolution of the convective radial transport of electron heat 
	for various RMP amplitudes.  For better visualisation, the transport is
	normalized to its maximum value at each time slice.
    }
    \label{fig:specnf}
\end{figure}

Fig.~\ref{fig:specnf} illustrates the time evolution of space-averaged
toroidal mode number spectra associated with the ELM-induced convective radial
transport of electron heat as defined by Eq.\ (\ref{eq:qee}). In order to make the
linear phase visible, the spectra are time-dependently normalized to their
maximum values. In the RMP-free case (\ref{fig:specnfa}),
the most unstable mode is the one with $n=7$.  For the lowest perturbation
amplitude $A_{0}=0.1$ (\ref{fig:specnfb}), the initial linear
phase for $t\lesssim15$, is superseded by the RMP-induced resonant mode number
$n=4$.  Later, for $t\gtrsim15$, the nominal mode with $n=7$ competes with a
resonant mode with $n=8$. For $A_{0}=1$ (\ref{fig:specnfc}), the
nominal mode is completely replaced by RMP-induced resonants, and for the largest
RM perturbation amplitude $A_{0}=10$ (\ref{fig:specnfd}), resonant
modes even prevail the nonlinear dynamics.

\begin{figure}[ht]
    \begin{center}
	\subfloat[\label{fig:qete}]{\includegraphics[width=71mm]{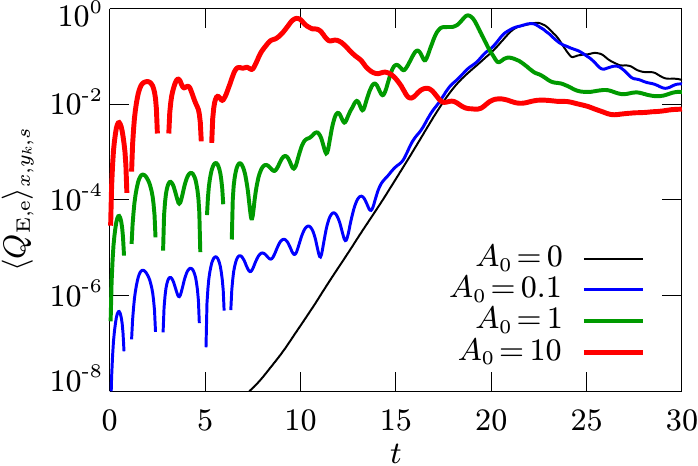}}\hfill
	\subfloat[\label{fig:qetm}]{\includegraphics[width=71mm]{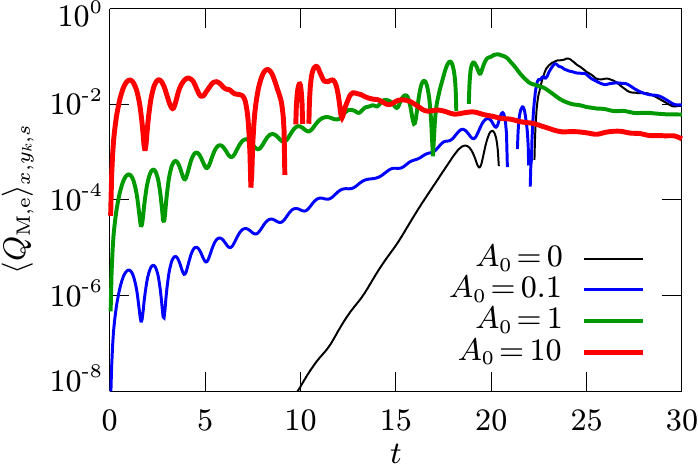}}
    \end{center}
    \caption{\sl
	Time variation of the volume-averaged (a) convective and (b) magnetic
	radial transport of electron heat for various RMP amplitudes. 
    }
    \label{fig:qet}
\end{figure}

Fig.~\ref{fig:specnfb} shows a situation where the nominal mode with
$n=7$ competes with the RMP-induced resonant mode with $n=8$. The interaction
between two competing neighbouring modes could have a mitigative effect on the
ideal ballooning ELM blow-out.  In Fig.\ \ref{fig:qet}, we investigate this
point by considering time traces of the volume-averaged convective and magnetic
radial transport of electron heat as given by Eqs.\ (\ref{eq:qee},\ref{eq:qem}). 
The maximum of the convective heat transport (Fig.\ \ref{fig:qete}) is shifted
from $t\approx22$ in the perturbation-free case to $t\approx10$ for the highest
perturbation amplitude $A_{0}=10$. Moreover, the amplitude of the peak
transport is slightly increased with increasing perturbation amplitude. The
magnetic transport (Fig.\ \ref{fig:qetm}) exhibits similar characteristics:
RMPs cause a shift of the peak transport to earlier times and give rise to a
significant increase of the magnetic transport in the linear phase. For the
perturbation-free case, the magnetic contribution to the transport in the
linear phase is negligible.  By contrast, magnetic and convective transport
become comparable if a magnetic perturbation is applied. As the
RMP-free linear ballooning instability substantially preserves the
magnetic flux surfaces (the associated magnetic flutter emerges only in the
nonlinear phase), this effect can be ascribed to
the RMP-induced formation of ergodic field regions which enhance the magnetic
transport in the linear phase.

A mitigation of the ideal ballooning ELM blow-out due to a competition between
the most unstable mode associated with the RMP-free case and the
RMP-induced resonant modes is not observed. On the contrary, RMPs involve the
formation of stationary density fluctuations \cite{reiser05}, which drive the
growth of a resonant MHD component. This can be explained by the fact that the 
amplitudes of the quasi-turbulent fluctuations associated with the initially
prescribed pedestal state are small so that the RMP-induced resonant structures
can easily prevail.  If the initial, quasi-turbulent density fluctuations are increased
by a factor of 1000, the linear growth phase of the nominal mode with $n=7$ is
shortened so that this mode is dominant up to perturbation amplitudes
$A_{0}\leq1$. For $A_{0}=10$, we again observe a competition between the modes
with $n=7$ and $n=8$. However, a mitigation of the ideal ballooning ELM
blow-out is not observed.

The above results indicate that the present ideal ballooning unstable pedestal
profile scenario is not appropriate to investigate a possible RMP-induced
mitigation effect on ELMs.

\subsection{Summary and discussion of results on RMP suppression of edge localized bursts}

The effects of RMPs on ideal ballooning unstable edge profiles were examined.
An ``H-mode'' model scenario described in \cite{kendl10,peer13} was considered. 

The main findings can be summarized as follows:
\begin{enumerate}
    \item Within the ideal ballooning unstable pedestal scenario, the RMP-induced
	formation of resonant structures drives the formation of ballooning
	blow-outs which are resonant with the magnetic perturbation field.
    \item RMPs cause a modification of the zonal profiles of electron density
	and electric potential by up to 30~$\%$. The temporal
	fluctuations of the zonal profiles are strongly reduced.
    \item The examined ``H-mode''-like model scenarios do not allow to simulate a mitigation
	of ideal ballooning ELMs by RMPs.
\end{enumerate}

The above simulations constitute a first approach to study the effects of RMPs
on ELM-like bursts within the gyrofluid electromagnetic model GEMR. 
A direct mitigation of edge loclaised ideal ballooning modes could not be
established in the simulations. On the contrary, RMPs were found to drive the
formation of ideal ballooning ELMs.  
In Ref.\ \cite{canik10} a destabilization of ELMs by RMPs in otherwise
quiescent H-mode states has actually been reported.  
However, none of the proposed destabilization mechanisms is in agreement with
the RMP-induced formation of resonant structures which drive resonant IBM blow-outs.

\section{Conclusions}

We have considered the effects of externally-applied resonant magnetic
perturbations (RMPs) on tokakamak edge turbulence and ideal ballooning bursts
in 6-moment electromagnetic gyrofluid computations including zonal profile evolution.

The interpretation of the L-mode like turbulence simulation results leads to the
following physical picture.  
RMPs give rise to plasma response currents which screen the vacuum
perturbation fields.  Even for strongly ergodized vacuum perturbation fields,
the amplitude of the intrinsic magnetic flutter and the resulting intrinsic
ergodicity are hardly changed by RMPs. As a consequence, the radial transport
by parallel motion along radially perturbed magnetic field lines is not
increased. Even for strongly ergodic vacuum perturbation fields, the radial
transport is mainly due to fluid-like $E\times B$ convection.  As the plasma
adjusts to the RMP-induced three-dimensional magnetic equilibrium, the
convective transport is reorganized in terms of convection cells which are
resonant with the toroidal and poloidal mode numbers of the perturbation
fields.  Accordingly, the radial convective transport exhibits a stationary
component which increases with increasing RMP amplitude. For the same reason,
RMPs give rise to a decrease of the turbulent fluctuations in the thermal state
variables, whereas the stationary contributions to the fluctuations increase.
Moreover, the RMP-induced stationary structures decelerate the poloidal plasma
rotation. Except for the imposed stationary structures, the drift-wave mode
structure of the turbulent fluctuations is widely preserved.
 
The computations faced the problem that the artificial dissipation slightly
varied with the RMP amplitude. Consequently, the total (convective and
magnetic) transport was not completely preserved across the simulations. The
problem was increased by an additional boundary dissipation which was found to
be necessary for stable RMP simulations. The main findings (screening of RMPs
by plasma response currents, dominance of the convective over the magnetic
transport) are not affected by this constraint. By contrast, a comparison
between the absolute values of the fluctuations in the dependent variables has
to be interpreted carefully. Nevertheless, the tendency for an RMP-induced
decrease of the fluctuations is clear.

If RMPs are applied to an ideal ballooning unstable ``H-mode''-like initial
pedestal configuration, the profiles adjusts to the RMP-induced three-dimensional
magnetic equilibrium and form resonant perturbations in the dependent variables. 
The interchange-ballooning drive increases the resonant perturbations so that
they can grow faster than the most unstable mode of the perturbation-free
case. 

The presented simulations constitute a first gyrofluid approach towards a
simulation of RMP effects on edge turbulence and on ELM mitigation. 
A direct mitigation of ideal ballooning mode bursts in modelled
``H-mode''-like states could not be observed. 

In turbulence computations the RMP fields were found to considerably reduce
the pedestal profile gradients (to around a half for typical experimental
perturbation field strengths). Following this picture, ELMs are likely
suppressed if in an experimental H-mode scenario the edge profiles are
sustained below a ballooning unstable critical gradient by RMP effects on the
(inter-ELM) turbulent transport.

\section*{Acknowledgements}

This work was mainly supported by the Austrian Science Fund (FWF) Y398. 
This work has been carried out within the framework of the EUROfusion
Consortium and has received funding from the Euratom research and training
programme 2014-2018 under grant agreement No 633053. The views and opinions
expressed herein do not necessarily reflect those of the European Commission.

\end{document}